\documentclass[preprint,nofootinbib,showpacs,preprintnumbers,amsmath,amssymb]{revtex4}

\usepackage{graphicx}
\usepackage{dcolumn}
\usepackage{bm}
\usepackage{slashed}
\newcommand{\bc}{\begin{center}}
\newcommand{\ec}{\end{center}}
\newcommand{\be}{\begin{equation}}
\newcommand{\ee}{\end{equation}}
\newcommand{\ba}{\begin{eqnarray}}
\newcommand{\ea}{\end{eqnarray}}
\newcommand{\bea}{\begin{eqnarray*}}
\newcommand{\eea}{\end{eqnarray*}}
\usepackage{orcidlink}

\PassOptionsToPackage{
    linktocpage=true,
    pdfproducer=medialab,
    pdfa=true
}{hyperref}

\begin{document}

\title{Accounting for correlations in the $\omega\pi$ spectral function and its implication on the leading model dependent contribution to the $\tau^- \to \pi^-\pi^0 \nu_\tau\gamma$ decay}

\author{Leonardo Esparza-Arellano\,~\orcidlink{0009-0007-9282-6131}}
\affiliation{Instituto de F\'{\i}sica,  Universidad Nacional Aut\'onoma de M\'exico, AP 20-364, Ciudad de M\'exico 01000, M\'exico}%
\author{Fabiola Fortuna\,~\orcidlink{0000-0002-8938-7613}}
\affiliation{Instituto de F\'{\i}sica,  Universidad Nacional Aut\'onoma de M\'exico, AP 20-364, Ciudad de M\'exico 01000, M\'exico}%
\author{Antonio Rojas\,~\orcidlink{0000-0002-8195-9517}}
\affiliation{Instituto de F\'{\i}sica,  Universidad Nacional Aut\'onoma de M\'exico, AP 20-364, Ciudad de M\'exico 01000, M\'exico}%
\author{Genaro Toledo\,~\orcidlink{0000-0003-4824-9983}}
\affiliation{Instituto de F\'{\i}sica,  Universidad Nacional Aut\'onoma de M\'exico, AP 20-364, Ciudad de M\'exico 01000, M\'exico}%

\date{\today}

\begin{abstract}
The $\omega\pi$ spectral function, measured in the decay $\tau^- \to \pi^-\omega\nu_\tau$ by the CLEO Collaboration, is described in the vector meson dominance approach. We sample the parameter space using a Metropolis Monte-Carlo algorithm and generate a set of parameters of the corresponding form factor, explicitly accounting for the correlation among them. 
We show the improvement in the uncertainty region of the spectral function with respect to neglecting correlations. We test the conserved vector current hypothesis by showing the agreement of the spectral function with the expected from the measurement of the cross section of $e^+e^- \to \omega\pi$ by the SND experiment, and the consistency of the branching ratio of the $\tau^- \to \pi^-\omega\nu_\tau$ decay  with the value reported by the Particle Data Group.
As an application, we consider the $\omega$-channel of the $\tau^- \to \pi^-\pi^0\nu_\tau\gamma$ process, the leading model dependent contribution in the vector meson dominance approach. We compute the contribution to the long distance electromagnetic correction function $G_\text{EM}(t)$, of the pure $\omega$-channel and its interference with the structure independent part. We show that the uncertainties, using the set of correlated parameters, are largely reduced with respect to neglecting correlations. The implication on the isospin symmetry breaking correction to the muon magnetic dipole moment estimation from tau data is found to be consistent with previous results but with improved precision. 
\end{abstract}

\maketitle

\section{Introduction}
 The $\omega\pi$ spectral function obtained from the $\tau^- \to \pi^-\omega\nu_\tau$ decay has proven to be a useful way to explore different aspects of the hadronic current and its implications on precision observables \cite{Tsai:1971vv, Decker:1987mn,Fajfer:1992hi,LopezCastro:1996xh,Flores-Tlalpa:2007hbv,Davier:2005xq,Guo:2008sh,Volkov:2012gv,Chen:2022nxm,Chen:2024xqt}. It has been measured by the ARGUS \cite{ARGUS:1986bzw} and CLEO \cite{CLEO:1987ftz, CLEO:1999heg} Collaborations, where a general analysis provided information on the possible underlying dynamics associated with its behavior. 
A proper theoretical description requires considering not only the value of the parameters involved, with the corresponding uncertainty, but also the correlation between them. This last becomes an issue when using the spectral function information in other processes. For example, the theoretical description of related observables, such as the branching ratio of the $\tau^- \to \pi^-\omega\nu_\tau$ decay, can be made artificially compatible with the spectral function data considering the parameters as uncorrelated, which enlarges the uncertainties. More elaborated relations, such as the one from the conserved vector current (CVC) hypothesis, allow testing its equivalence with the measurements from $e^+ e^- \to \pi\omega$ \cite{Eidelman:1997zt,Eidelman:2001ju,Eidelman:2001ee,Davier:2002dy}, in this case deviations might be wrongly associated to isospin symmetry breaking corrections.
Thus, a theoretical description of the measured spectral function that also provides an accurate and consistent description of other observables can be considered as a reliable representation of the data. 
Having such a description of the measured spectral function allows us to explore its effect on the theoretical estimate of the muon magnetic dipole moment (MDM). The leading hadronic vacuum polarization (HVP) and the Light-by-Light scattering contributions require, to some extent, proper knowledge of the $\omega \pi$ form factor to provide an accurate theoretical prediction \cite{Alemany:1997tn, Aliberti:2025beg, Davier:2002dy}. Here, we are concerned with the decay $\tau^- \to \pi^-\pi^0\nu_\tau\gamma$, 
which enters in the long distance electromagnetic factor $G_\text{EM}(t)$, where $t$ is the square of the two pions invariant mass, to account for the corresponding isospin symmetry breaking correction of the HVP estimate using tau data.
In the vector meson dominance approach (VMD), the main model dependent (MD) contribution of this process can be traced to the channel involving an intermediate $\omega$ meson, which then decays into a pion and a photon \cite{Flores-Tlalpa:2005msx,FloresBaez:2006gf,FloresTlalpa:2006gs}, this is the so-called $\omega$ channel. It involves a double-pole enhancement mechanism \cite{Esparza-Arellano:2023dps}, making its contribution important regardless of its higher order in the photon energy expansion, with respect to the model-independent part~\cite{CENPLB,CEN,Flores-Tlalpa:2005msx,FloresBaez:2006gf,FloresTlalpa:2006gs,LopezCastro:2015cja,pablo20,Masjuan:2023qsp, Chen:2022nxm}.
The main source of uncertainty for this contribution is associated with a lack of accurate information on the parameters involved and the correlations between them \cite{Flores-Tlalpa:2005msx,FloresBaez:2006gf,FloresTlalpa:2006gs,Esparza-Arellano:2023dps,avalos}. 
Similar difficulties are faced in the Chiral description at $\mathcal{O}(p^6)$, where analogous contributions appear \cite{pablo20}. Thus, an accurate treatment of this channel is essential for determining the isospin-symmetry-breaking electromagnetic correction to the muon MDM from tau data. Here, we follow previous works \cite{Flores-Tlalpa:2005msx,FloresBaez:2006gf,FloresTlalpa:2006gs} for the general description in the VMD approach. 
Contributions are of two types: interference with the other channels and the pure $\omega$ channel. 
In the analysis of the \(\tau^-\to\pi^-\pi^0\nu_\tau\) decay data by the Belle Collaboration \cite{Belle:2008xpe} they removed the radiative part coming from the pure \(\omega\) contribution, accounting for a branching ratio of $0.1 \pm 0.01$ \%. However, the Particle Data Group (PDG) quotes $\text{Br}(\tau \to \omega\pi\nu_\tau)= 1.95 \pm 0.06 \ (\%)$ and $\text{Br}(\omega\to \pi\gamma)= 8.33 \pm 0.25 \ (\%)$ \cite{PDG26}. Then, in a narrow width approximation the radiative process must be close to $0.16\pm 0.01 \ (\%)$
This suggests that the reported subtraction may not fully account for this contribution. Therefore it is important to assess the potential impact of a residual pure \(\omega\) component on the muon anomalous magnetic moment. The SI–\(\omega\) interference contribution is not removed from data and therefore must be fully considered. A comparison of our results with the corresponding estimates in the literature will be drawn.

Other effects such as the proper matching between the long and short distance electromagnetic isospin breaking contributions and the hadronic structure effects are also relevant, but collateral to the scope of this work~\cite{Aliberti:2025beg, Colangelo:2025iad,Colangelo:2025ivq,Cirigliano:2026ios,Allen:2026iad}.\\
Here, we aim to describe the $\omega\pi$ spectral function obtained from the $\tau^- \to \pi^-\omega\nu_\tau$ decay, and study its implication in several observables.
In Section \ref{sec:fit}, we present the model for the spectral function in the VMD approach. Then we fit the data obtained by the CLEO Collaboration \cite{CLEO:1999heg} using MINUIT, which is used as a reference, and sample the parameter space using a Markov-chain-Monte-Carlo (MC) based on the Metropolis algorithm. With the latter, we generate a set of configurations of the parameters, explicitly accounting for the correlation between them. We present our description of the spectral function and quantify the resulting uncertainty band, emphasizing the role of correlations among the fit parameters.
In Section \ref{sec:test}
we test the CVC hypothesis, showing the agreement with the spectral function expected from the measurement of the \(e^+e^- \to \omega\pi\) cross section by the SND experiment \cite{snd2pg16}. We also obtain the branching ratio of the decay $\tau^- \to \pi^-\omega\nu_\tau$, which is found to be consistent with the value reported by the PDG.
In Section~\ref{sec:omega-channel}, as an application, we consider the $\omega$-channel of the  $\tau^- \to \pi^-\pi^0\nu_\tau\gamma$ process. We compute the branching ratio of the pure $\omega$ channel and determine the fraction corresponding with the amount subtracted in the Belle analysis. We also evaluate the contribution to the electromagnetic correction function $G_\text{EM}(t)$ of the individual channel and its interference with the structure independent part. We obtain the uncertainties by evaluating them in the set of correlated parameters, and compare with previous estimates. In Section~\ref{sec:g-2}, we discuss the implications for the electromagnetic correction to the muon anomalous magnetic dipole moment. Finally, we present our conclusions in Section~\ref{sec:conclusions}.

\section{The $\omega\pi$ spectral function}\label{sec:fit}

The spectral function obtained from the $\tau^- \to \pi^-\omega\nu_\tau$ process has been studied in several approaches \cite{Tsai:1971vv, Fajfer:1992hi,LopezCastro:1996xh,Flores-Tlalpa:2007hbv,Davier:2005xq,Guo:2008sh,Volkov:2012gv,Chen:2022nxm,Chen:2024xqt} and was measured by ARGUS \cite{ARGUS:1986bzw} and CLEO \cite{CLEO:1987ftz, CLEO:1999heg} Collaborations. 
The general form is \cite{Tsai:1971vv,ARGUS:1986bzw,Davier:2005xq}
\begin{equation}
    v(s)=\frac{32 \pi^2 m_\tau^3}{G_F^2 V_{ud}^2(m_\tau^2-s)^2(m_\tau^2+2s)}\frac{d\Gamma}{ds},
\label{eq:spectral1}
\end{equation}
where $d\Gamma/ds$ is the differential decay width and $s$ is the square of the $\omega\pi$ invariant mass. In order to describe the process in the VMD approach, we follow Ref.~\cite{LopezCastro:1996xh}.
 We set the notation as $\tau^-(p) \to \omega(q_1, \epsilon^*) \pi^-(q_2) \nu_\tau(q)$, where the quantities in parentheses denote the corresponding four-momenta, and $\epsilon^*$ is the omega polarization vector. The decay amplitude is
\begin{equation}
    \mathcal{M}=\frac{i \ G_F V_{ud}^* \ l^\mu}{\sqrt{2}}F_{\omega\pi}(s) \epsilon_{\mu\alpha\beta\gamma} \epsilon^{*\alpha}q_1^\beta q_2^\gamma,
\label{eq:amplitude}
\end{equation}
where $s=(q_1+q_2)^2$ and $l^\mu=\bar{u}(q)\gamma^{\mu}(1-\gamma_5)u(p)$ is the leptonic tensor. The form factor $F_{\omega\pi}(s)$ is modeled considering the resonances $\rho(770)$ and $\rho(1450)$, denoted by $\rho$ and $\rho^\prime$, respectively \footnote{CLEO analysis shows that the inclusion of the resonance $\rho(1700)$ (denoted by $\rho^{\prime\prime}$) does not further improve the description of the data \cite{CLEO:1999heg}.}, as
\begin{equation} \label{eq:form_factor}
    F_{\omega\pi}(s)=\frac{G_\rho \ g_{\rho\omega\pi}}{m_\rho^2-s-i\sqrt{s}\Gamma_\rho(s)}+\frac{e^{i\theta} \ G_{\rho^\prime} \ g_{{\rho^\prime}\omega\pi}}{m_{\rho^\prime}^2-s-i \sqrt{s}\Gamma_{\rho^\prime}},
\end{equation}
where $G_V$ and $g_{V\omega\pi}$ ($V=\rho,\rho^\prime$) are the corresponding couplings of the vector state to the weak current and to the $\omega\pi$ mesons, respectively, and $\theta$ is the relative phase between them.
Using Eq.~(\ref{eq:amplitude}), the differential decay width is 
\begin{equation}
    \frac{d\Gamma}{ds}=\frac{G_f^2 V_{ud}^2}{1536 \pi^3 m_\tau^3}|F_{\omega\pi}(s)|^2 \frac{(m_\tau^2-s)^2(m_\tau^2+2s)}{s^2} \lambda(s,m_\omega^2,m_\pi^2)^{3/2},
    \label{eq:dgds}
\end{equation}
where $\lambda(x,y,z)$ is the K\"allén function. Thus, the spectral function, in terms of the form factor is
\begin{equation}
    v(s)=|F_{\omega\pi}(s)|^2 \frac{\lambda(s,m_\omega^2,m_\pi^2)^{3/2}}{48 \pi s^2}.
\label{spectralVMD}\end{equation}
Note that CLEO \cite{CLEO:1999heg} uses the same definition for the spectral function as in Eq.  (\ref{eq:spectral1}), but its analytical expression in terms of the form factor is not the same. 
We use the KSFR relation $ G_\rho=\sqrt{2}m_\rho^2/g_{\rho\pi\pi}$ \cite{Kawarabayashi:1966kd,Riazuddin:1966sw} to fix the weak coupling of the $\rho$ meson, where $g_{\rho\pi\pi}=5.949 \pm 0.076$ \cite{avalos} is the strong coupling between $\rho$ and two pions. The form factor can be set as 
\begin{equation}
    F_{\omega\pi}(s)= \frac{\sqrt{2} \ g_{\rho\omega\pi}}{g_{\rho\pi\pi}}
    f_{\omega\pi}(s),
    \label{eq:Fomegapi}
    \end{equation}
    enforcing the normalization condition
    \begin{equation}
    f_{\omega\pi}(s)=\frac{1}{1+B^\prime e^{i\theta}}\Big[ BW_\rho(s)+B^\prime e^{i\theta}  BW_{\rho^\prime}(s)
    \Big],
\end{equation}
where $BW_V(s)\equiv m_V^2/( m_V^2-s-i\sqrt{s}\Gamma_V)$, using the corresponding energy-dependent width, as discussed below, and $B^\prime=(G_{\rho^\prime} \ g_{{\rho^\prime}\omega\pi} \ m_\rho^2)/(G_{\rho} \ g_{{\rho}\omega\pi}  \ m_{\rho^\prime}^2)\equiv B/g_{{\rho}\omega\pi}$. Thus, it has three free parameters: the coupling $g_{\rho\omega\pi}$, the $\rho^\prime$ to $\rho$ relative magnitude $B$, and the relative phase $\theta$. We consider the mass and total decay width of the resonances as given by the PDG \cite{PDG26}. For $\rho$ we consider an energy-dependent width, with contributions from the $\pi\pi$, $KK$ and $\omega\pi$ channels
\begin{equation}
    \Gamma_\rho(s)=\Gamma_{\rho\to \pi\pi}(s)\Theta(s-4m_\pi^2)+\Gamma_{\rho\to KK}(s)\Theta(s-4m_K^2)+\Gamma_{\rho\to\omega\pi}(s)\Theta(s-(m_\omega+m_\pi)^2),
\end{equation}
where
\begin{eqnarray}
    \Gamma_{\rho\to \pi\pi}(s)&=&\Gamma_\rho \frac{m_\rho^5}{s^{5/2}}\,\left[\frac{\lambda(s,m_\pi^2,m_\pi^2)}{\lambda(m_\rho^2,m_\pi^2,m_\pi^2)}\right]^{3/2},\\
    \Gamma_{\rho\to KK}(s)&=&\frac{\Gamma_\rho}{2} \frac{m_\rho^5}{s^{5/2}}\,\left[\frac{\lambda(s,m_K^2,m_K^2)}{\lambda(m_\rho^2,m_\pi^2,m_\pi^2)}\right]^{3/2},\\
        \Gamma_{\rho \to\omega\pi}(s)&=&\frac{g_{\rho\omega\pi}^2}{96 \pi s^{3/2}}\lambda(s,m_\omega^2,m_\pi^2)^{3/2},
\end{eqnarray}
with the relation $g_{\rho KK}=g_{\rho \pi\pi}/2$ from $SU(3)$ flavor symmetry. For the \(\rho'\), we assume a constant decay width.\\ 
In a first stage, we fit the spectral function data measured by CLEO \cite{CLEO:1999heg} using MINUIT~\cite{James:1975dr}, as a reference. Since the full experimental covariance matrix is not available, bin-to-bin correlations are neglected.
In Table \ref{parameters1} second row, we show the result for the three free parameters and the $\chi^2/dof$ of the fit. The correlation between parameters can be quantified globally through the Pearson correlation coefficient. We show them in Table \ref{tab:correlations} as obtained from MINUIT. 
In addition, a more detailed characterization of the spectral function, suitable for evaluating different observables and their associated uncertainties, can be obtained through MC sampling using the Metropolis algorithm. This procedure generates a set of configurations that accounts for correlations between the fit parameters, allowing the observables and their uncertainties to be  statistically evaluated. Thus, in a second stage, we describe the data using a standard Metropolis Monte-Carlo algorithm (MC) \cite{Metropolis:1953am} with the normal probability distribution function given by
\begin{equation}
    P(x)= e^{-\chi^2/2}
\end{equation}
with
\begin{equation} \label{eq:chi2}
    \chi^2= \sum_i \frac{(v(g_{\rho\omega\pi}, B, \theta)_i- v_{\text{exp}_i})^2}{\sigma_i^2}
\end{equation}
where $i$ runs over the experimental data points ~\cite{Piekarewicz:2014kza}. We consider the restriction conditions $B>0$ and $-\pi<\theta<\pi$. We initially generated 5000 configurations to thermalize the system. Then, we generated 50,000 configurations from which a corresponding set of 50,000 triads of correlated parameters was obtained \footnote{To test the convergence, an analysis considering 3000, 5000, and 10000 configurations for thermalization led to similar results. Statistical analysis using blocks of every 5 and 10 steps  gives similar results. 
This set allows us to determine statistical quantities such as the average ($\bar{x}$), the standard deviation ($\sigma_x$), and the correlation coefficient ($\rho_{xy}$).}
\begin{eqnarray}
    \bar{x}&=& \frac{1}{N}\sum_{i=1}^N x_i,\\
    \sigma_x^2&=& \frac{1}{N}\sum_{i=1}^N (x_i-\bar{x})^2,\\
    \rho_{xy}&=&\frac{1}{N}\sum_{i=1}^N \frac{(x_i-\bar{x})(y_i-\bar{y})}{\sigma_x \sigma_y}.
\end{eqnarray}
For non-Gaussian distributions we use the median, and for the interval of a one standard deviation, the quantiles that give the same probability content \cite{PDG26}.
In Table \ref{parameters1} third row, we show the result for the parameters, obtained using the MC procedure to describe the spectral function. The quoted $\chi^2/dof$ corresponds to the evaluation of Eq.~(\ref{eq:chi2}) at the the median values of the parameters. 
The value of the parameters obtained must be interpreted collectively, as they are obtained from a description of the spectral function, which involves a correlation between them. Consequently, comparing individual parameter values directly with other determinations in the literature may be misleading. In particular, our value of \(g_{\rho\omega\pi}\) can not be directly compared with the uncorrelated determination in Ref.~\cite{avalos}.

\begin{table}[t]
\begin{center}
\begin{tabular}{|c|c|c|c|c|}
     \hline
     \hline
     Set&$g_{\rho\omega\pi}$ (GeV$^{-1})$ & $B$ (GeV$^{-1})$ & $\theta$ (rad) &$\chi^2/dof$ \\ 
     \hline 
     \hline
MINUIT  & $26.13^{+4.99}_{-5.33}$ & $3.00^{+0.18}_{-0.17}$ & $1.27^{+0.48}_{-0.35}$ & 1.14\\
    \hline
\hline
MC & $24.92_{-6.56}^{+5.73}$ & $3.02^{+0.19}_{-0.17}$ & $1.38_{-0.41}^{+0.82}$ & 1.28\\
\hline
\hline
\end{tabular}
\end{center}
\caption{Parameters of the model to describe the $\omega\pi$ spectral function, obtained  from the fit using MINUIT and the result from Monte-Carlo.}
\label{parameters1}
\end{table}


\begin{table}
    \centering
    \begin{tabular}{|c|ccc|}
       \hline
       &$g_{\rho\omega\pi}$ &  $B$& $\theta$ \\
       \hline 
         $g_{\rho\omega\pi}$ &  1  & 0.379 & -0.818\\       
         $B$&0.379 & 1 & -0.357\\        
         $\theta$& -0.818 & -0.357 & 1\\
         \hline
    \end{tabular}
    \caption{Correlation coefficients from MINUIT.} 
    \label{tab:correlations}
\end{table}

\begin{figure}
\centering
\includegraphics[scale=0.7,angle=0]{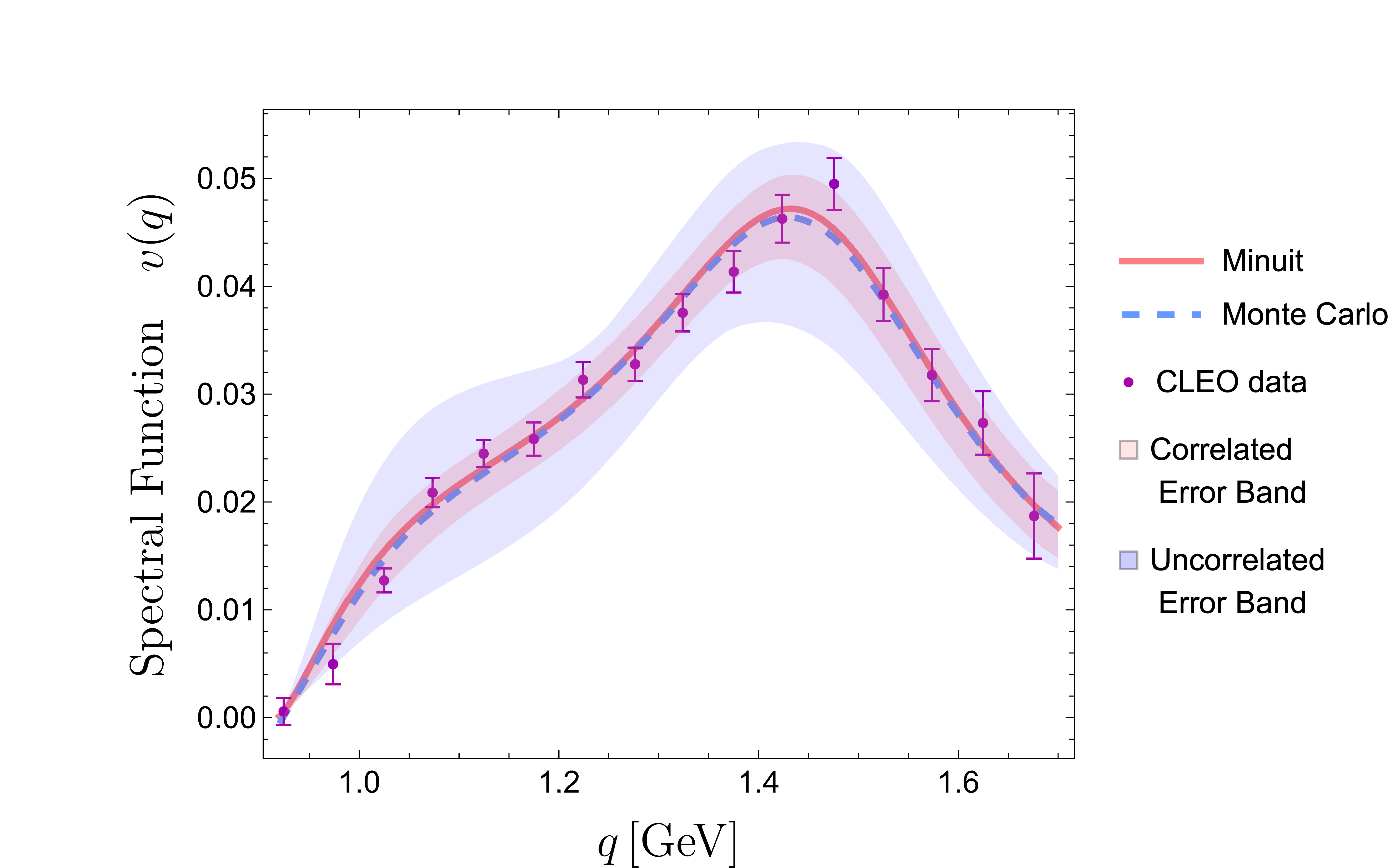}
\caption{$v(s)$ spectral function as a function of $q=\sqrt{s}$. CLEO data (symbols), fit using MINUIT (solid line) and MC (dashed line). The uncertainty regions correspond to uncorrelated/correlated (larger/smaller shaded area) parameters.\\
}
\label{fig:Minuit}
\end{figure}

In Fig. \ref{fig:Minuit}, we show the spectral function from CLEO data (symbols with error bars), the results considering the central values from MINUIT (solid line) and from MC (dashed line). In the same figure, we compare the one-standard-deviation correlated error band, obtained from MC (smaller shaded band), with the band generated from the uncorrelated case (larger shaded band). 
The latter is generated by considering the point-to-point largest variation of the spectral function for all combinations of positive and negative uncertainties shifts of the parameters obtained by MINUIT ~\footnote{Not a single combination of parameters produces the upper or lower delimiting curves.}. 
The correlated case is obtained by evaluating the spectral function in the set of parameters and determining the corresponding standard deviation. We observe a significant reduction of the error band when considering the correlations with respect to the uncorrelated scenario.

\begin{figure}[t]
\centering
\includegraphics[scale=0.5]{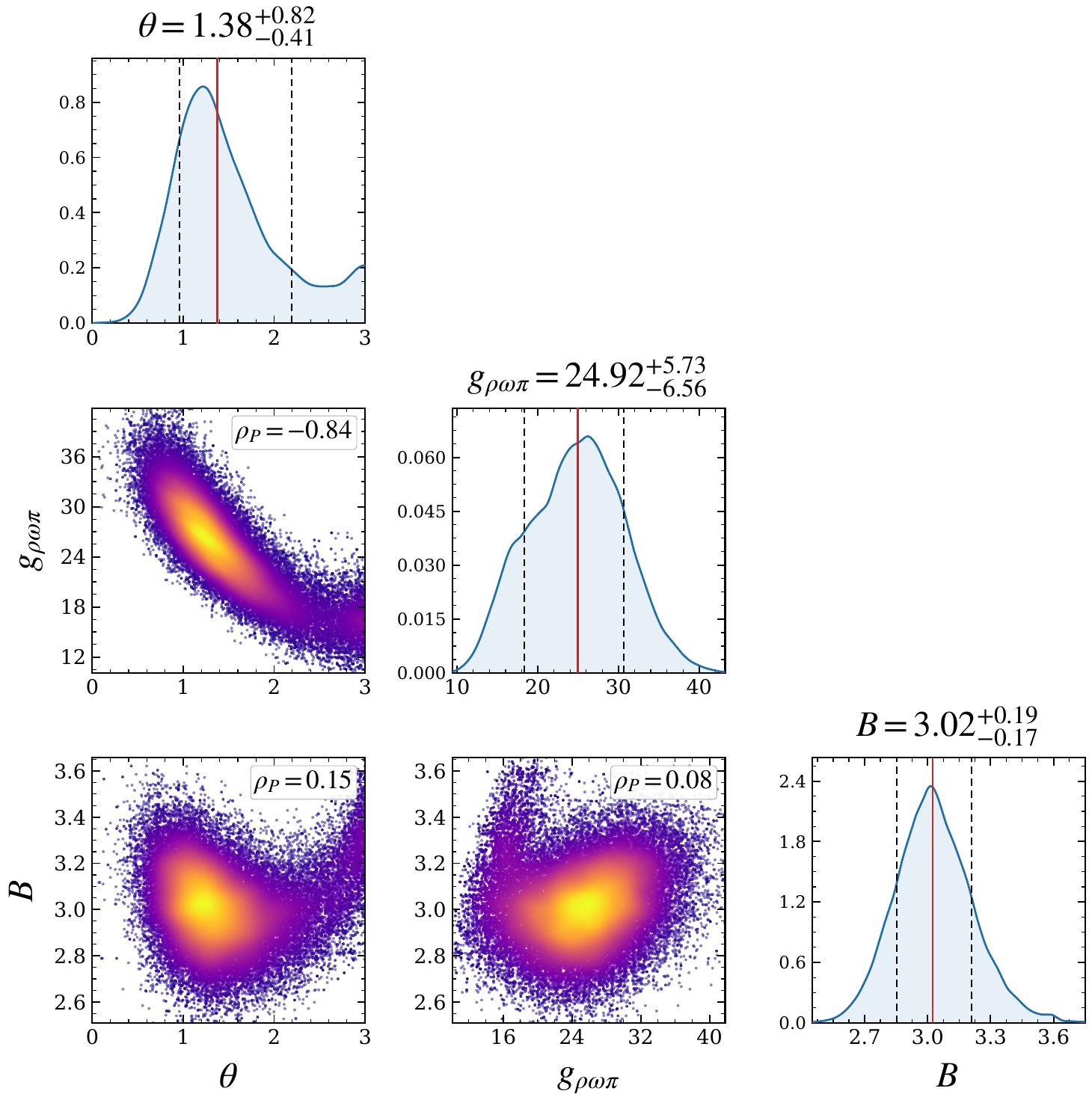}
\caption{Corner plot. Histogram of $\theta$(rad), $g_{\rho\omega\pi}$(GeV$^{-1}$), and $B$(GeV$^{-1}$) obtained using MC and their correlated distribution $g_{\rho\omega\pi}-B$, $B-\theta$, and $g_{\rho\omega\pi}-\theta$. $\rho_{P}$ corresponds to the Pearson correlation coefficient between the corresponding parameters in $x$ and $y$ axis.}
\label{fig:corrAll}
\end{figure}

In Fig. \ref{fig:corrAll}, we show the corner plot, including the histograms of the parameters as obtained using MC, and the sampled joint distribution between the correlated parameters $g_{\rho\omega\pi}-B$, $g_{\rho\omega\pi}-\theta$, and $B-\theta$. 
The correlation between the fitted parameters was first characterized by the Pearson correlation coefficient ($\rho_P$) which provides a compact measure of linear association. Since the sampled joint distributions are not strictly elliptical, we additionally evaluated the Spearman rank correlation ($\rho_S$) and the distance correlation $(\mathrm{dCor})$ as robustness diagnostics \cite{Schober:2018,Szekely:2007}. The resulting values for \((g_{\rho\omega\pi},\theta)\), \((B,\theta)\), and \((B,g_{\rho\omega\pi})\) are \(\rho_P=(-0.84,0.15,0.08)\), \(\rho_S=(-0.86,-0.03,0.10)\), and \(\mathrm{dCor}=(0.84,0.24,0.16)\), respectively. The three diagnostics lead to the same qualitative hierarchy: the dominant dependence occurs between \(g_{\rho\omega\pi}\) and \(\theta\), whereas the correlations involving \(B\) are considerably weaker. We therefore quote the Pearson coefficients in Fig. \ref{fig:corrAll} as compact summary statistics, while the full non-Gaussian dependence structure is displayed directly through the sampled joint distributions and their highest-density regions.
While the dominant \(g_{\rho\omega\pi}-\theta\) anticorrelation is consistently identified by both MINUIT and MC, weaker correlations involving \(B\) differ between the local covariance estimate and the sampled global distribution. This behavior reflects the non-elliptical structure visible in Fig. \ref{fig:corrAll} and motivates propagating the sampled joint distribution rather than relying exclusively on the local covariance matrix.
We observe that accounting for the correlations reduces the range over which the parameters can significantly vary.

\section{CVC hypothesis and branching ratio}
\label{sec:test}
\begin{figure}[t]
\centering
\includegraphics[width=13cm]{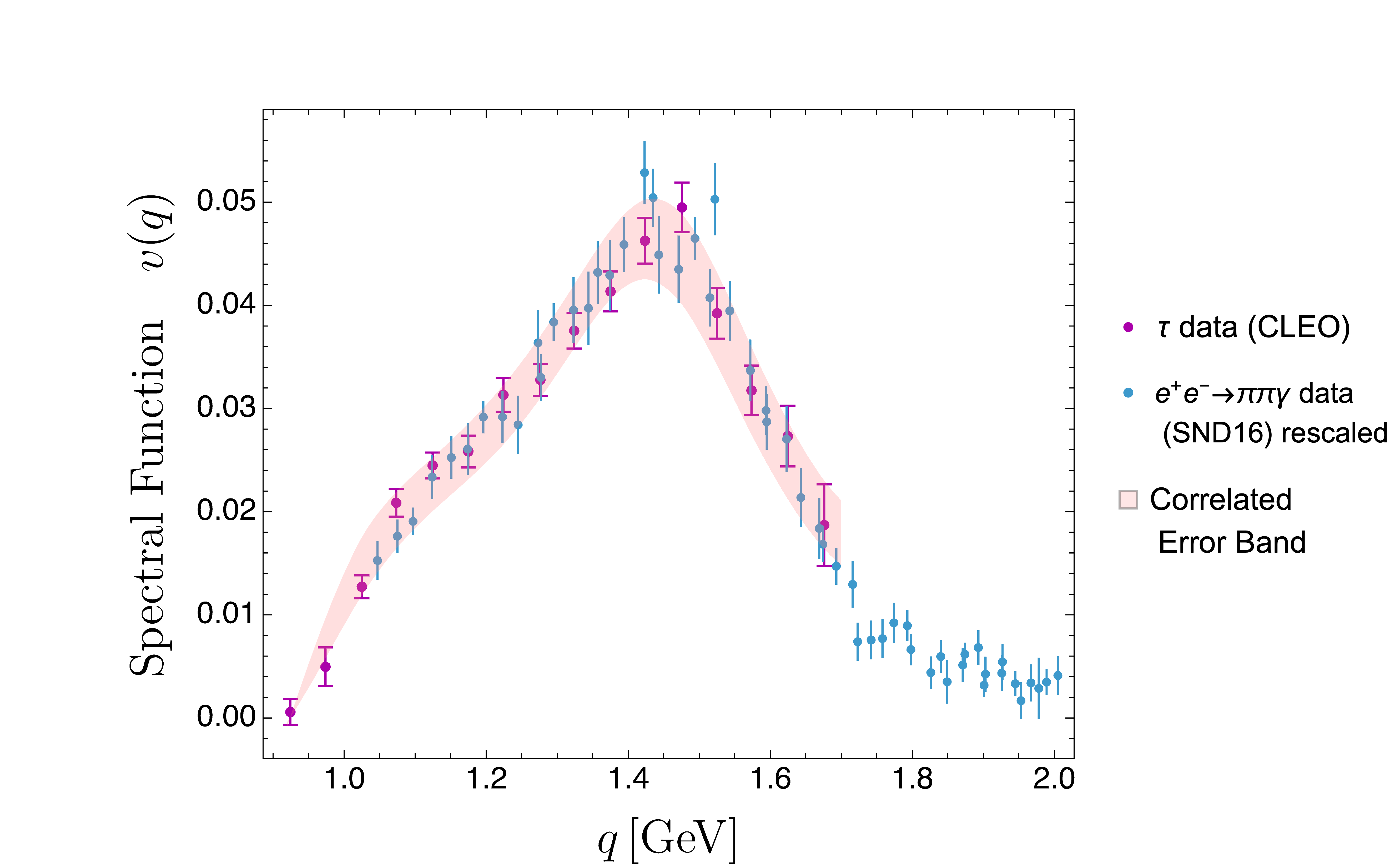}
\caption{\label{fig:CVC} CVC test of the $v(s)$ spectral function.  CLEO data from $\tau \to\nu_\tau \pi\omega$ (magenta symbols), SND data from $e^+ e^- \to \pi\omega$ (blue symbols) and MC error band (shaded area).}
\end{figure}

\begin{figure}[t]
\centering
\includegraphics[width=10cm]{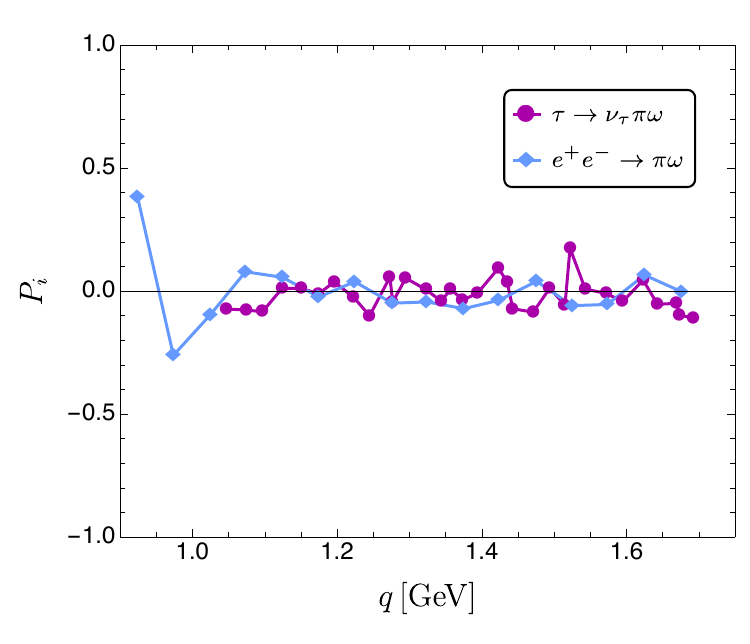}
\caption{\label{fig:pull} CVC test. The Pull function $P_i$ comparing the spectral function data from CLEO $\tau \to\nu_\tau \pi\omega$ (circle symbols) and SND $e^+ e^- \to \pi\omega$  (diamond symbols) relative to the MC description of the CLEO data.}
\end{figure}

The CVC hypothesis allows us to relate the $e^+ e^- \to \textit{hadrons}$ cross section to the $\tau \to \nu_\tau\, \textit{hadrons}$ decay \cite{Eidelman:1997zt,Eidelman:2001ju,Eidelman:2001ee,Davier:2002dy}. In particular, the spectral function extracted from $\tau \to \omega\pi\nu_\tau$, measured by CLEO, can be related to the corresponding $e^+e^- \to \pi^0\omega$ cross section by \cite{Tsai:1971vv}
\begin{equation}
    v^{\tau}(q^2)= \frac{q^2 \sigma^{e^+ e^-}_{I=1}(q^2)}{4 \pi^2 \alpha^2}.
\end{equation}
The experimental data for the cross section of the $e^+e^- \to \pi^0\omega$ process can be obtained by rescaling the $e^+e^- \to \pi^0\omega \to \pi^0\pi^0 \gamma$ cross section by the $\omega \to \pi^0 \gamma$ branching ratio \cite{PDG26}. Using the measurement performed by SND \cite{snd2pg16} we have obtained the corresponding CVC spectral function. In Fig. \ref{fig:CVC}  we plot it together with the CLEO spectral function to exhibit the consistency between both descriptions. We also plot the correlated uncertainty region, observing that it covers both data sets within the uncertainties of the corresponding individual data sets, confirming the CVC hypothesis at the current precision.\\
An estimator between two data sets $y^{(1,2)}$, with their corresponding standard deviations $\sigma_{1,2}$, is the Pull function defined as
\begin{equation}
    P_i\equiv\frac{y_i^{(1)}-y_i^{(2)}}{\sqrt{\sigma_{1,i}^2+\sigma_{2,i}^2}},
\end{equation}
where $i$ is the index for the corresponding energy $q_i$ at which they are obtained. 
Figure~\ref{fig:pull} shows the Pull distributions for two comparisons: CLEO (circle symbols) and SND (diamond symbols) spectral-function data are each compared with the MC description of the CLEO data. We restricted the evaluation around the common kinematical region. We observe that both the CLEO and SND data are consistent with the MC description and, therefore, also with each other.

The branching ratio and the corresponding uncertainty for the decay $\tau \to \omega\pi\nu_\tau$ can be computed by the integration of Eq. (\ref{eq:dgds}) evaluated in the set of correlated parameters, for which we obtain

\begin{equation}
    \text{Br}(\tau \to \omega\pi\nu_\tau)= 2.07 \pm 0.15 \ (\%).
    \label{eq:Brnr}
\end{equation}
Following a similar procedure, we obtain the corresponding result for the CVC hypothesis
\begin{equation}
    \text{Br}(\tau \to \omega\pi\nu_\tau)_\text{CVC}= 2.03 \pm 0.19 \ (\%),
  \label{eq:Brnr}
\end{equation}
where we have described the SND data, using MC, in the energy range up to tau mass, having a $\chi^2/dof=1.87$ for the median of the parameters.
These results are consistent with the value reported by the PDG, $\text{Br}(\tau \to \omega\pi\nu_\tau)= 1.95 \pm 0.06$\ (\%) \cite{PDG26}. The PDG result relies partially on the experimental information by CLEO, while the prediction from CVC hypothesis is an independent result. It is interesting to observe that they all have a similar precision level.

\section{$\omega$ channel of $\tau^- \to \pi^-\pi^0 \nu_\tau\gamma$ decay} \label{sec:omega-channel}
We now explore the implications of our description of the spectral function in the radiative decay $ \tau^{-}(p) \rightarrow \pi^{-} \ (p_-) \ \pi^{0} \ (p_0) \ \nu_{\tau} (q)\gamma \ (k, \epsilon^*)$, where in parentheses are the corresponding
four-momenta and $\epsilon^*$ is the polarization vector of the photon. We follow the same approach as in Refs.~\cite{Flores-Tlalpa:2005msx,FloresBaez:2006gf,FloresTlalpa:2006gs}. 
The most general form of the total amplitude for the process can be written as \cite{CEN,Bijnens:1992en}:
\begin{eqnarray}
{\cal{M}}_{T}&=&eG_FV^*_{ud}\epsilon^{*\mu}\Big[F_{\nu}\bar{u}(q)\gamma^{\nu}(1-\gamma_5)(m_{\tau}+\slashed p-\slashed k)\gamma_{\mu}u(p)\nonumber\\
&&\hspace{2 cm}+(V_{\mu\nu}-A_{\mu\nu})\bar{u}(q)\gamma^{\nu}(1-\gamma_5)u(p)\Big],\label{TCEN}
\end{eqnarray}
where $G_F$ is the Fermi constant and $V^*_{ud}$ is the corresponding Cabibbo-Kobayashi-Maskawa matrix element.
The first line in Eq.~(\ref{TCEN}) describes the $\tau$ radiation, where $F_{\nu} \equiv Q_\nu\frac{f_+[t]}{2\,p\cdot k}$, $f_+[t]$ is the pion form factor, $Q \equiv p_{0}-p_{-}$ and $t$ is the square of the two pion invariant mass.
The $V_{\mu\nu}$ and $A_{\mu\nu}$ tensors correspond to the Vector and Axial contributions of the $W^{*-}\rightarrow \pi^-\pi^0\gamma$ transition, respectively. $V_{\mu\nu}$ has the following structure:
\begin{eqnarray}
 V_{\mu\nu} &=&-f_+[t^{\prime}]\frac{p_{-\mu}}{p_-\!\cdot\! k}(Q-k)_{\nu}-f_+[t^{\prime}]g_{\mu\nu}\nonumber\\
&&+\frac{f_+[t^{\prime}]-f_+[t]}{k\!\cdot\! (p_{-}+p_0)}(p_{-}+p_0)_{\mu}Q_{\nu} + \hat{V}_{\mu\nu},
\end{eqnarray}
where
\begin{eqnarray}
  \hat{V}_{\mu\nu} &\equiv&v_1\,p_-\!\cdot\! k\,F_{\mu\nu}(p_-) + v_2\, p_0\!\cdot\! k\, F_{\mu\nu}(p_0)\nonumber\\
 &&+v_3\,p_0\!\cdot\! k\,\,p_-\!\cdot\! k\,L_{\mu}(p_-,p_0)\,p_{-\nu} + v_4\,p_0\!\cdot\! k\,\,p_-\!\cdot\! k L_{\mu}(p_-,p_0)(p_{-}+p_0+k)_{\nu},\label{Vmunu}
\end{eqnarray}
and we have defined the variable $t^{\prime}=t+2(p_{-}+p_0)\!\cdot\! k$, and the tensor functions
$
L_{\mu}(a,b) \equiv a_{\mu}/a\cdot k-b_{\mu}/b\cdot k
$,
$F_{\mu\nu}(a) \equiv g_{\mu\nu} -a_{\mu}k_{\nu}/a\cdot k$. The functions $v_i$  $(i=1,..,4)$ encode the particular form of the vector contribution.
We will consider the amplitude at the structure independent level (SI), where we neglect the axial contribution $A_{\mu\nu}=0$, which  is very small compared to the vector part \cite{CEN,FloresBaez:2006gf,FloresTlalpa:2006gs}. In this approach, the $v_i$ functions receive contributions from the magnetic dipole moment of the $\rho$ meson  (taken as 2 in units of $e/2m_\rho$). The only non-zero terms are \cite{LopezCastro:1999dp} 
\begin{equation}
     v_1 = -v_2 = \frac{[f_+(t^{\prime})-f_+(t)]}{(p_{-}+p_0)\!\cdot\! k}.\nonumber\\ 
 \label{vis}
\end{equation}
We can identify the Low amplitude ${\cal{M}}_{Low}$, as the part that includes the contributions of order  $O(k^{-1})$ and $O(k^{0})$ \cite{Low}. This is  the model independent part (MI), whose explicit form can be found elsewhere \cite{CENPLB,CEN,FloresBaez:2006gf,FloresTlalpa:2006gs, pablo20}.
The form factor $f_+(t)$ was extracted from a fit to the two pion invariant mass distribution of the $\tau \to \nu_\tau\pi\pi$ decay, measured by Belle \cite{Belle:2008xpe}. It includes the $\rho$, $\rho^\prime$ and $\rho^{\prime\prime}$ vector mesons. The explicit form and the parameters can be found in Ref.~\cite{Esparza-Arellano:2023dps}.

\begin{figure}
\centering
\includegraphics[scale=0.5]{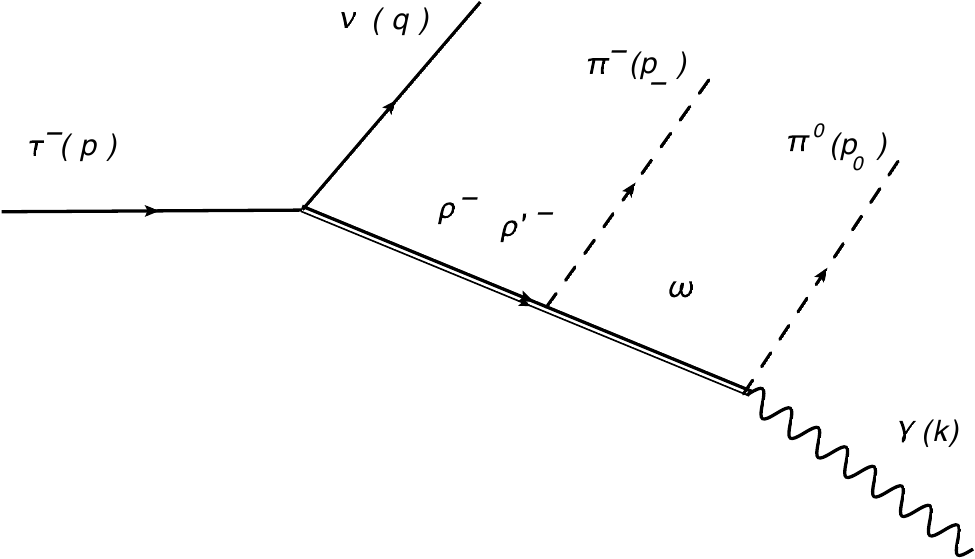}
\caption{\centering $\omega$-channel of the decay $\tau^- \to \pi^-\pi^0 \nu_\tau \gamma$.}
\label{Fig:omega}
\end{figure}

In the VMD approach, the only relevant model dependent (MD) channel is the one with an $\omega$ exchange \cite{FloresBaez:2006gf,FloresTlalpa:2006gs}, depicted in Fig. \ref{Fig:omega}. This is the decay $\tau^- \to \pi^-\omega \nu_\tau$ previously studied, with the subsequent decay of the $\omega$ meson to $\pi^0 \gamma$, in the four-body kinematical region. The amplitude, contributing to $\hat{V}_{\mu\nu}$, is
\be
{\cal M}_\omega=eG_F V^*_{ud} \frac{\hat{g}_{\rho\omega\pi}}{\sqrt{2}\, g_\rho m_\rho^2 m_\omega^2} BW_\omega(r)
F_{\omega\pi}(t^\prime)
{\epsilon_{\alpha\sigma\mu}}^\lambda
{\epsilon_{\phi\lambda\chi\nu}}
k^\sigma p_0^\alpha (p_0+k)^\phi p_-^\chi \epsilon^{*\mu}  \ell^{\nu},
 \label{omega}
\ee
where $r\equiv (p_0+k)^2$. The form factor $F_{\omega\pi}(t^\prime)$ is the one previously obtained from the spectral function of the decay $\tau^- \to \pi^- \omega \nu_\tau$, Eq.~(\ref{eq:Fomegapi}), evaluated at $t^\prime$. The coupling $\hat{g}_{\rho\omega\pi}=11.314\pm 0.383$ GeV$^{-1}$ \cite{avalos} is associated with the decay $\omega \to \pi^0\gamma$. We distinguish the numerical value of this parameter from that involved in the spectral function, where it was treated as a free but correlated parameter, intended to describe such an observable. 
The contribution of the $\omega$ channel to the process enters through the square of the corresponding amplitude (pure $\omega$), and its interference with the SI part (SI--$\omega$ interference). \\
Using the amplitude given in Eq.~(\ref{omega}), we compute the contribution of the pure $\omega$ channel to the branching ratio of the $\tau^-\to \pi^-\pi^0\nu_\tau\gamma$ decay. The four-body kinematics is implemented following Ref.~\cite{Flores-Tlalpa:2005msx}. The corresponding central value and uncertainty are obtained by evaluating the decay rate over the set of configurations of the parameters, resulting in the following
\begin{equation}
    \text{Br}(\tau^-\to \pi^-\pi^0\nu_\tau\gamma)_\omega = 0.16 \pm 0.02 \ (\%). 
    \label{eq:Bromegach}
\end{equation}
This result can be compared with the fraction of the omega channel removed as background from the data in the study of the $\tau^-\to \pi^-\pi^0\nu_\tau$ process by Belle \cite{Belle:2008xpe}. They quote $\tau \to \nu_\tau \pi \omega$ with $\omega \to \pi\gamma$ of $0.10 \pm 0.01$ (\%), indicating that in our description the pure $\omega$ channel is larger than the one removed by Belle.
As reference, the PDG gives $0.16\pm 0.01 \ (\%)$, as mentioned before, which is similar to Eq.~(\ref{eq:Bromegach}).
 We have determined that the Branching ratio, evaluated in an energy region of $(m_\omega -\Gamma_\omega,m_\omega +\Gamma_\omega)$ gives a value consistent with the central value reported by Belle. We will refer to the outside region as the off--\(\omega\) part.\\

The photon inclusive decay $\tau^-\to \pi^-\pi^0\nu_\tau(\gamma)$  takes into account both virtual and real photon emission. In this case, the di-pion invariant mass distribution can be related to the corresponding non-radiative decay distribution, $d\Gamma^0_{2\pi}/dt$, through the long-distance electromagnetic correction function, $G_{\rm EM}(t)$, as
\begin{equation}
\frac{d\Gamma_{2\pi(\gamma)}}{dt} = \frac{d\Gamma^0_{2\pi}}{dt}G_\text{EM}(t)\,.
\end{equation}
This function is usually divided into two parts \cite{CENPLB,CEN,FloresBaez:2006gf}
\begin{equation}
G_\text{EM}(t) = G^0_\text{EM}(t) + G^\text{rest}_\text{EM}(t)\,,
\label{Gemtsplit}
\end{equation}
$G^0_\text{EM}(t)$ accounts for the virtual and real photon emission up to $O(k^{-2})$ and $G^\text{rest}_\text{EM}(t)$ includes the remaining MI and MD higher order contributions from the real photon emission. An early calculation of $G^0_\text{EM}(t)$ was made in Refs.~\cite{CENPLB,CEN}, and more recently in Refs.~\cite{Colangelo:2025iad, Colangelo:2025ivq, Cirigliano:2026ios} it was reanalyzed incorporating the pion structure in the loops and the matching to the short distance electroweak correction factor $S_\text{EW}$. Since our study is concerned only with the MD contribution to real-photon emission, for comparison purposes we split it as
\begin{equation}
G_\text{EM}(t) = G^\text{SI}_\text{EM}(t) + G^{\omega}_\text{EM}(t)\,,
\label{Gemtsplit}
\end{equation}
where the SI part was computed in Refs.~\cite{Flores-Tlalpa:2005msx,FloresBaez:2006gf,FloresTlalpa:2006gs, Esparza-Arellano:2023dps} and includes \(G_{\rm EM}^{0}(t)\), as obtained in Refs.~\cite{CENPLB,CEN} and discussed above, while the \(\omega\)-channel contribution comprises both the SI–\(\omega\) interference and the pure \(\omega\)-channel.

\begin{figure}
\includegraphics[width=10cm]{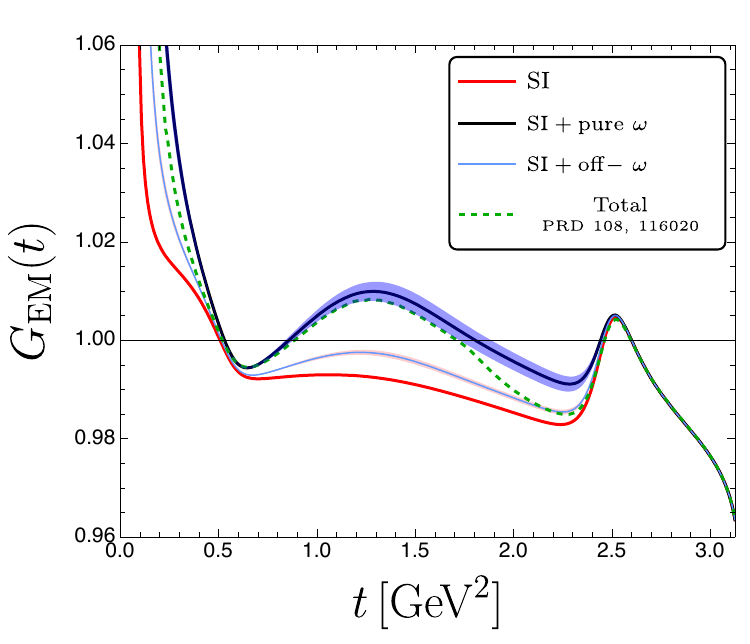}
\caption{$G_\text{EM}(t)$ contributions: SI (red solid line), SI plus the pure $\omega$ (black line), and the SI plus the off--$\omega$ (blue line). The corresponding shaded region represents the $1\sigma$ uncertainty. The green dashed line shows a previous result for the total $G_\text{EM}(t)$ (including the SI-$\omega$ interference and the pure $\omega$ channel)~\cite{Esparza-Arellano:2023dps}.}
\label{Gemto}
\end{figure}

\begin{figure}
\includegraphics[width=10cm]{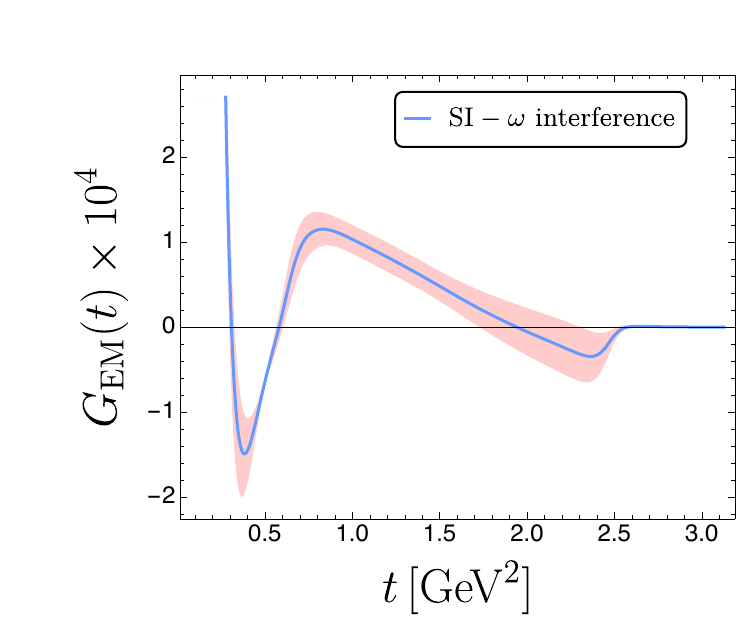}
\caption{$G_\text{EM}(t)$ function including only the interference between SI contribution and the $\omega$ channel. The shaded pink region represents the 1$\sigma$ uncertainty.}
\label{Gemtint}
\end{figure}

 In Fig. \ref{Gemto}, we show this electromagnetic function including different contributions: the SI contribution (red solid line), the sum of SI and the pure $\omega$ channel (black solid line), and the sum of SI and the off-$\omega$ channel (blue solid line); the corresponding shaded region refers to the uncertainty obtained by adding in quadrature the uncertainty produced by the correlated parameters and that from $\hat{g}_{\rho\omega\pi}$.
 We also plot the central value for the total process (green dashed line) obtained in Ref.~\cite{Esparza-Arellano:2023dps}, but omit its large uncertainty region.
Given the difference in scales between these contributions and the SI--$\omega$ interference, including the latter does not produce visible changes. In Fig. \ref{Gemtint}, we show only the SI--$\omega$ interference contribution and the corresponding uncertainty region under the same considerations. Notice that while remaining at the $\mathcal{O}(10^{-4})$ level over most of the $t$ range, it rises rapidly near the threshold.

\section{Correction to the $\tau$-based muon MDM estimate} \label{sec:g-2}

The leading hadronic contribution, $a_\mu^\text{(HVP,\,LO)}$ to the muon MDM, based on $\tau$ data, comes from the decay into two pions. It requires to account for the correction, $\Delta a_\mu^\text{(HVP,\,LO)}$, due to all the isospin symmetry breaking sources, $R_\text{IB}(t)$. In addition to $G_\text{EM}(t)$ and $S_\text{EW}$~\cite{Erler:2002mv}, it involves the final state radiation from the pions, the phase space modification and the correction accounting for the difference between the form factors describing the charged-current \(\tau\) decay and the neutral-current \(e^+e^-\) production processes. These corrections have been computed, with the main source of uncertainty coming from the form factors ratio and the $G_\text{EM}(t)$ term \cite{Aoyama:2020ynm,CENPLB,CEN,Davier:2010nc,pablo20,Davier:2010fmf,Jegerlehner:2017gek, Benayoun:2012etq,Benayoun:2021ody,Aliberti:2025beg,Flores-Baez:2025nra, Allen:2026iad, Bruno:2026yba}. 
Incorporating the hadronic structure in radiative corrections of the difference between the neutral and charged $\rho$-meson widths, as well as in the final-state radiation, is also part of the effort to reduce the uncertainties \cite{Flores-Baez:2025nra}. Here, we focus on the correction $\Delta a_\mu^\text{(HVP,\,LO)}$ denoted as $\Delta a_\mu^{{[\pi\pi,\tau]}}$ from $G_\text{EM}(t)$. This can be estimated by \cite{ Davier:2010fmf}
\begin{eqnarray}\label{delta-amu}
\Delta a_\mu^{{[\pi\pi,\tau]}}\vert_{G_\text{EM}(t)}&=&\frac{\alpha^2 m_\tau^2}{6 |V_{ud}|^2\pi^2}\frac{\mathcal{B}_{\pi\pi^0}}{\mathcal{B}_e}\int^{m_\tau^2}_{4m^2_{\pi}} dt \, \frac{K(t)}{t}
\left(1-\frac{t}{m_\tau^2}
\right)^{-2}
\left(1+\frac{2t}{m_\tau^2}
\right)^{-1} \nonumber\\
&&\frac{dN_{\pi\pi^0}}{N_{\pi\pi^0}dt}
\times \Big[\frac{1}{G_\text{EM}(t)}-1\Big],
\end{eqnarray}
where $K(t)$ is the QED Kernel function~\cite{Brodsky:1967sr}, $(1/N_{\pi\pi^0})dN_{\pi\pi^0}/ds$ is the normalized invariant mass spectrum of the $\pi\pi^0$ final state and $\mathcal{B}_{X^-}$ denotes the branching fraction of $\tau^-\to X^-(\gamma)\nu_\tau$. We use ${\cal B}_{\pi\pi^0} =(25.42\pm 0.09)\%$ \cite{Davier:2010fmf}, ${\cal B}_e=(17.82\pm 0.04)\%$, and $|V_{ud}|=0.97367(32)$ \cite{PDG26}. For the normalized spectrum of two-pions in tau decays we use a combined set of ALEPH-CLEO-OPAL-Belle data \cite{private, Davier:2010fmf}. 

Using the central value of $G_{\rm EM}(t)$ and its uncertainty region, we can compute the corresponding contributions.
In Table \ref{amutabla}, we show the results for $\Delta a_\mu^{[\pi\pi,\tau]}\vert_{G_\text{EM}(t)}$ considering several contributions. Upper panel: SI, SI--$\omega$ interference, pure $\omega$, and off-$\omega$. The uncertainties include the statistical from the MC generated data set, and a second uncertainty of 15\% associated with the modeling, intended to cover variations in the form factors parameterization, $g_{\rho\pi\pi}$ coupling uncertainty, and missing effects associated with the energy dependence of the $\rho^\prime$ width. We expect no effects from $SU(3)$ flavor symmetry breaking on the $g_{\rho KK}$ coupling, as this channel is subdominant in the $\rho$ decay width.

For comparison, in the intermediate panel we show the SI plus SI--$\omega$ interference, along with the results for the same contribution reported by the $g$-2 Working Group in the White Paper 2025 (WP25)~\cite{Aliberti:2025beg}, Davier et al. \cite{Davier:2010fmf}, Miranda et al.~\cite{pablo20}, Castro et al. \cite{Castro:2024prg}, and Li et al.~\cite{Li:2025zus}. We observe that the central values are consistent between them (except for~\cite{Li:2025zus}), but the uncertainties obtained in this work are significantly smaller. Note that the uncertainty quoted in Ref.~\cite{Davier:2010fmf} is an overestimation intended to cover the difference between their result and the one obtained in Refs. \cite{CENPLB,CEN}, rather than a model uncertainty.
In the last row, we quote the result from Esparza et al.~\cite{Esparza-Arellano:2023dps}, which includes the sum of the SI, SI--$\omega$ interference, and pure $\omega$ contributions.

Adding the isospin symmetry breaking contribution from phase space and $S_\text{EW}$ as given in Ref.~\cite{Aliberti:2025beg} to our result for $G_\text{EM}(t)$ (SI plus SI-$\omega$ interference) we obtain: $(-22.0\pm 0.2) \times 10^{-10}$. This value is consistent with the combined result reported in the same reference, $(-21.9\pm 1.9)\times 10^{-10}$.
This can also be compared with the result by Cirigiliano et al.~\cite{Cirigliano:2026ios} of $(-24.9\pm 0.5)\times 10^{-10}$ and Colangelo et al.~\cite{Colangelo:2025iad} of $(-24.8\pm 0.5)\times 10^{-10}$. As pointed out in Ref.~\cite{Colangelo:2025iad}, the difference in the treatment of the pion structure and the matching with the short-distance part shifts the central value by about 3 units. The off--$\omega$ contribution may also shift the overall result by $-3\pm 0.6$.

\begin{table}
\begin{tabular}[t]{|c|c|}
\hline
 $G_\text{EM}(t)$  & $\Delta a_\mu^{[\pi\pi,\tau]}$  ($ \times 10^{-11}$)  \\
\hline
SI & $-13.8$  \\
SI--$\omega$ interference & $-4.8 \pm 0.9 \pm 0.7 $ \\
pure $\omega$  & $-95.8 \pm {10.9} \pm 14.4$ \\
 off-$\omega$  & $-31.0 \pm 3.9\pm 4.7$ \\
 \hline
  SI + interference & $-18.6\pm 0.9 \pm 0.7$ \\
  WP25~\cite{Aliberti:2025beg} & $-20\pm 14$\\
  Davier et al.~\cite{Davier:2010fmf} & $-19.2\pm 9$\\
  Miranda et al.~\cite{pablo20}&$-15.9^{+5.7}_{-16}$\\
  Castro et al.~\cite{Castro:2024prg} & $-16.7^{+6.0}_{-13.9}$\\
  Li et al.~\cite{Li:2025zus} & $-6.3 \,\,/  -44.4$\\
  \hline
 Total, Esparza et al.~\cite{Esparza-Arellano:2023dps} & $-94.03^{+32.2}_{-92.0}$ \\
\hline
\end{tabular}
\caption{ $\Delta a_\mu^{[\pi\pi,\tau]}|_{G_\text{EM}(t)}$  ($ \times 10^{-11}$) for several contributions of $G_\text{EM}(t)$: Upper panel: SI, SI--$\omega$ interference, pure $\omega$, and off-$\omega$ obtained in this work. The intermediate panel shows the  SI + interference and a set of results in the literature. Last row is the SI plus SI\,-\,$\omega$ interference and pure omega obtained in Ref.~\cite{Esparza-Arellano:2023dps}. See text for details.}
\label{amutabla}
\end{table}

\section{Conclusions} \label{sec:conclusions}

We have described the $\omega\pi$ spectral function data measured by the CLEO Collaboration in the decay $\tau^- \to \pi^-\omega\nu_\tau$, within the vector meson dominance approach. We sampled the parameter space using a Metropolis Monte-Carlo algorithm, explicitly accounting for the correlation between them. This procedure generated a set of parameter configurations to be used in the evaluation of different observables involving the information carried out by the spectral function. The parameters were characterized based in their statistical distributed properties. We exhibited the benefit of this approach by observing the reduction on the uncertainty region describing the spectral function with respect to neglecting correlations. We tested the CVC hypothesis by comparing the CLEO measurement of the spectral function with the one inferred from the $e^+e^- \to\pi^0\pi^0\gamma$ cross-section measurement by the SND experiment, finding good agreement between them within the current uncertainties, quantified using a Pull function and the branching ratio of the decay $\tau^- \to \pi^-\omega\nu_\tau$. The latter was also found to be consistent with the value reported by the PDG.
As an application of the description of the spectral function, we considered the $\omega$-channel of the $\tau^- \to \pi^-\pi^0\nu_\tau\gamma$ process. 
The branching ratio from this particular channel was found to be consistent with the estimate using the PDG branching ratio of the sub-processes but larger than the one reported by the Belle Coll. for the fraction subtracted in the analysis of the $\tau^- \to \pi^-\pi^0\nu_\tau$ decay. We defined an off-$\omega$ contribution as the remaining part after subtracting the fraction removed by Belle.
We computed the contributions to the long-distance electromagnetic correction function $G_{\rm EM}(t)$ of the pure $\omega$ channel, its interference with the SI part, and the off-$\omega$ region. We determined their corresponding uncertainty region. The dominant pure $\omega$ uncertainty region can be compared with previous estimates for the total contribution, observing that it is substantially smaller.
We then computed the isospin symmetry-breaking correction to the muon magnetic dipole moment estimate from $\tau$ data arising from these contributions. \\
Since our analysis focused exclusively on the model-dependent contribution, we combined it with the previously calculated structure-independent part. The resulting estimate is consistent with previous determinations but has significantly smaller uncertainties.
A comparison of the combined effect of the phase space, \(S_{\text{EW}}\), and \(G_{\text{EM}}(t)\) shows that our result is consistent with previous estimates that do not account for either the matching between short- and long-distance contributions or the pion structure. It is also consistent with estimates that incorporate these refinements, once the corresponding shift is taken into account.
A possible remaining contribution from the off--$\omega$ part was identified and may need further investigation.\\
Our systematic description of the spectral function, incorporating the correlation between parameters, proved to be a useful ingredient in improving the uncertainties of observables, paving the way for more robust predictions.

\begin{acknowledgments}
We thank G. L\'opez Castro, P. Roig, Alejandro Miranda and J. Piekarewicz for very useful observations.
We acknowledge the support of SECIHTI, Mexico Grant No. 1104553 (L.E.A.) and the support of DGAPA-PAPIIT UNAM, under Grant No. IN105526. This research was supported by the Munich Institute for Astro-, Particle and BioPhysics (MIAPbP) which is funded by the Deutsche Forschungsgemeinschaft (DFG, German Research Foundation) under Germany´s Excellence Strategy – EXC-2094 – 390783311. The work of F.F. is funded by \textit{Estancias Posdoctorales por México, Estancia Posdoctoral Iniciales, Secihti}.
\end{acknowledgments}

\bibliographystyle{unsrt}
\bibliography{omegabiblio.bib}
\end{document}